\pdfoutput=1
\RequirePackage{filecontents}

\documentclass{article}
\providecommand{\keywords}[1]{\textbf{\textit{Index terms---}} #1}

\usepackage[top=1in, bottom=0.75in, left=0.75in, right=0.75in]{geometry}
\usepackage{fixltx2e}
\usepackage{graphicx}
\usepackage{subfig}
\usepackage{array,cite}
\usepackage{amssymb}
\usepackage{mathrsfs}
\usepackage[cmex10]{amsmath}
\graphicspath{{./Figures/}}
\newtheorem{theorem}{Theorem}%[section]

\newtheorem{assumption}{A.}
\newtheorem{remark}{Remark}
\newtheorem{lemma}{Lemma}

\title{\LARGE \bf
Internal Model Based Active Disturbance Rejection Control
}

\author{Jinwen Pan and Yong Wang*% <-this % stops a space
\thanks{The authors are all with Department of Automation, University of Science and Technology of China, Hefei 230027, P.R. China. (Corresponding author's e-mail: {\tt\small yongwang@ustc.edu.cn})}}

\begin{document}
%\bstctlcite{IEEEexample:BSTcontrol}

\maketitle
\thispagestyle{empty}
\pagestyle{empty}

%%%%%%%%%%%%%%%%%%%%%%%%%%%%%%%%%%%%%%%%%%%%%%%%%%%%%%%%%%%%%%%%%%%%%%%%%%%%%%%%
\begin{abstract}
\footnote{Manuscript accepted for publication in the proceeding of the 2016 American Control Conference, July 6-8, Boston, MA, USA. \copyright 2016 IEEE. Personal use of this material is permitted. Permission from IEEE must be
	obtained for all other uses, in any current or future media, including
	reprinting/republishing this material for advertising or promotional purposes, creating new
	collective works, for resale or redistribution to servers or lists, or reuse of any copyrighted
	component of this work in other works.}
The basic active disturbance rejection control (BADRC) algorithm with only one order higher externed state observer (ESO) proves to be  robust to both internal and external disturbances. An advantage of BADRC is that in many applications it can achieve high disturbance attenuation level without requiring a detailed model of the plant or disturbance. However, this can be regarded as a disadvantage when the disturbance characteristic is known since the BADRC algorithm cannot exploit such information. This paper proposes an internal model based ADRC (IADRC) method, which can take advantage of knowing disturbance characteristic to achieve perfect estimation of the disturbance under some mild assumptions. The effectiveness of the proposed method is validated by comprehensive simulations and comparisons with the BADRC algorithm. 

\end{abstract}


%%%%%%%%%%%%%%%%%%%%%%%%%%%%%%%%%%%%%%%%%%%%%%%%%%%%%%%%%%%%%%%%%%%%%%%%%%%%%%%%

\section{Introduction}\label{introduction}
Rejecting unknown disturbances in dynamical systems is a fundamental control problem with various applications such as friction compensation during stick slip motion\cite{de1995new}, disturbance reduction in gyroscopes \cite{zhang2014adaptive,zhang2014adaptive2}, active noise control\cite{bodson2001active}, sinusoidal disturbances rejection of vibrating structures\cite{shahsavari2015adaptive,shahsavari2014adaptive}, control of robot manipulators\cite{horowitz1993learning}, rotating mechanisms control\cite{gentili2003robust}, and nano-positioning \cite{shahsavari2014repeatable,shahsavari2013limits}. This problem is usually solved by applying the internal model principle (IMP) for which a general solution is given in \cite{francis1975internal} in the case of linear systems. The IMP states that if the disturbance model can be accurately obtained and embedded in the controller, the disturbance can be entirely canceled. On the other hand, when there is no information available about the disturbance, IMP is no longer effective. 

Active disturbance rejection control (ADRC) was proposed by Han\cite{han2009from} as an alternative paradigm for control system design\cite{gao2001alternative,gao2006active}, and since it is a model--free approach, it has the inherent advantages of rejecting nonlinearities, uncertainties and disturbances. Fruitful simulation results as well as the experimental results have been reported in various applications\cite{dong2011active,dong2012robust,li2013active,tang2014development,pan2015flatness}. In these applications, the unknown parts (unknown nonlinearities, uncertainties and external disturbances) are treated as a total disturbance and estimated by an extended state observer (ESO). It has been proven that if the total disturbance or its first derivative is bounded, the estimate error is bounded and can be arbitrarily reduced\cite{yang2009capabilities}. However, the ESO has some limitations: (1) if the total disturbance is not a constant, the estimate error can only be bounded but not zero; (2) the disturbance information cannot be used. An example given in \cite{godbole2013performance} shows that perfect estimation of even a simple sinusoidal disturbance cannot be achieved by basic ESO.

In this paper, we will propose an internal model based active disturbance rejection control (IADRC) in consideration of the disturbance information. The disturbance is separated into two parts, which are the part that can be modeled and the part that cannot be modeled. The former part is estimated by a disturbance observer with estimate error exponentially converging to zero.  The unmodeled part with unknown nonlinearities and uncertainties are together treated as an extended state (total disturbance) of the system and estimated using ESO with a bounded error. It is shown that the modeled part is captured perfectly and the unmodeled part is regarded as a constant during the estimation and compensation from the IMP point of view. It is also illustrated that the performance of BADRC is improved significantly by IADRC when the disturbance information is used, and the more we know about the disturbance, the better IADRC performs. 

The remainder is organized as follows. The problem statement is described in section \ref{problemformulation}. In section \ref{disturbanceobserver}, a special class of disturbance that can be modeled as an output of a fully excited linear system is considered. Two adaptive estimation algorithms for the disturbance are proposed based on the known disturbance information. The IADRC is designed and analyzed in section \ref{IADRC}. Simulation examples are given in section \ref{simulationexamples}, and conclusions are drawn in section \ref{conclusions}.

\section{Problem formulation}\label{problemformulation}
Consider the nonlinear single-input-single-output system %, which can be transformed into the following form,
\begin{equation}\label{sys1}
\left\{ {\begin{array}{*{20}{l}}
	{\dot {\bar x} = \bar A\bar x + \bar f\left( {\bar x,{\omega _1}} \right) + \bar b\left( {u + {\omega _2}} \right)}\\
	{\bar y = \bar c\bar x}
	\end{array}} \right.,
\end{equation}
with the state vector $ \bar x = {\left[ {{x_1},{x_2}, \cdots ,{x_n}} \right]^{\rm{T}}} $, the control input $ u \in  R $, \textit{shift matrix} $ \bar A \in R^{ n \times n} $, $ f_n(\bar x,\omega _1) $ an entirely unknown nonlinear smooth function, and $ \bar f\left( {\bar x,{\omega _1}} \right) = {\left[ {0, \cdots ,0,{f_n}\left( {\bar x,{\omega _1}} \right)} \right]^{\rm{T}}} $. $ \bar b = {[0,\cdots ,0,{b_n}]^{\rm{T}}} $ with $ {b_n} $ a known constant, $ \omega _1, \omega_2 \in R $ are bounded unknown time-varying disturbances and $ \bar c = \left[ {1,0, \cdots ,0} \right] $. System \eqref{sys1} can be rewritten as
\begin{equation}\label{sys2}
\left\{ \begin{array}{l}
\dot x = Ax + b\left( {u + d_2} \right) + f\\
y = cx
\end{array} \right.,
\end{equation}
where $ A \in  R^{(n+1) \times (n+1)} $, $ b = {\left[ {{{\bar b}^{\rm{T}}},0} \right]^{\rm{T}}} $, $ c = \left[ {\bar c,0} \right] $, $ x = {\left[ {{{\bar x}^{\rm{T}}},{x_{n + 1}}} \right]^{\rm{T}}} $ and $ f = {\left[ {0,\cdots,0,{{\dot x}_{n + 1}}} \right]^{\rm{T}}} $ with the extended state $ x_{n+1} $ defined as
\begin{equation*}
{x_{n + 1}} = {f_n}\left( {\bar{x},{\omega _1}} \right) + {b_n}\left( {{\omega _2} - {d_2}} \right).
\end{equation*}
$ d_2 $ is part of the matched disturbance $ \omega _2 $ that has some known information. The total matched disturbance $ d $ is defined as
%\begin{equation}
$d = d_1+d_2$,
%\end{equation}
where 
%\begin{equation}
${d_1} := {{x_{n + 1}}}/{{b_n}}$,
%\end{equation}
and $ d_2 $ is the output of following system
\begin{equation}\label{exo}
\left\{ \begin{array}{l}
\dot w = Sw, \qquad w\left( 0 \right) = {w_0}\\
d_2 = {h^{\rm{T}}}w
\end{array} \right.,
\end{equation} 
with $ w \in R^{s} $ and $ w_0 $ is selected that $ w $ are fully excited.

\begin{remark}
The extended state $ x_{n+1} $ that is entirely unknown can be viewed as the lumped unknown disturbance consisting of unknown nonlinearities, uncertainties of the plant and unknown part of external disturbances. $ d_1 $ can be considered as part of the total matched disturbance that is entirely unknown.
\end{remark}

For the system and the disturbances, we have the following assumptions,
\begin{assumption}\label{A.1}
$ f_n(\bar x,\omega _1) $ is unknown, but $ \dot f_n(\bar x,\omega _1) $ or $ f_n(\bar x,\omega _1) $ is bounded with $ f_n(0,0) = C_1 $ where $ C_1 $ is an unknown constant.
\end{assumption}

\begin{assumption}\label{A.2}
The matrix $ S $ has no zero eigenvalues.
\end{assumption}

\begin{assumption}
$ h $ is an unknown constant vector.
\end{assumption}

\begin{assumption}
The matrix $ S $ is entirely known.
\end{assumption}

\begin{assumption}
The matrix $ S $ is unknown, but $ s $, the dimension of $ S $ is known.
\end{assumption}

Our problem is to design an output feedback controller to stabilize the origin with the ability to reject the disturbance $ d_2 $ exponentially (thus perfectly) when $ \omega _1 =0 $ by making full use of the known information of the external disturbance and simultaneously to reject $ d_1 $ in the frame of ADRC.

\section{Disturbance observer design and analysis}\label{disturbanceobserver}
The idea is as follows. we use an ESO to estimate the internal uncertainty and a disturbance observer to estimate the external disturbance exponentially, and then compensate the total disturbance. 

If the external disturbance does not exist in system  (\ref{sys1}), that is, $ d_2 = 0 $ in system  (\ref{sys2}). In this case, the extended state observer can be designed as
\begin{equation}\label{eso2}
\dot p = Ap + bu + l\left( {\bar y - cp} \right),
\end{equation}
where $ p,l \in R^{n+1} $ and $ l $ is chosen such that  $ A - lc $ is Hurwitz. It is difficult to estimate the real states because of the unknown disturbance $ d $. $ A - lc $ and $ S $ have exclusive eigenvalues for that we have assumption A.\ref{A.1} and the selection of $ l $, so unique solution $ Q \in R^{\left( {n + 1} \right) \times s } $ of the following Sylvester equation
\begin{equation}\label{Sylvester1}
QS = \left( {A - lc} \right)Q + b{h^{\rm{T}}},
\end{equation}
for a given $ S $ exists\cite{bartels1972solution}. By defining $q := Qw$, \eqref{Sylvester1} implies
\begin{equation}\label{filter1}
\dot q = \left( {A - lc} \right)q + bd_2.
\end{equation}

\begin{remark}
Since $ h $ is unknown, no matter whether $ S $ is known or not, the solution $ Q $ cannot be obtained from  (\ref{Sylvester1}), and the observer  (\ref{filter1}) is unimplementable for that $ d_2 $ is unknown. 
\end{remark}

In order to obtain $ p $ and $ q $, we have the following lemma \cite{ding2005adaptive}.

\begin{lemma}\label{L1}
The state variable $ x $ can be expressed as
\begin{equation*}
x = p + q + \varepsilon,
\end{equation*}
where $ p $ is from  (\ref{eso2}) with $ q $ from  (\ref{filter1}) and $ \varepsilon $ satisfying
\begin{equation}\label{vare}
\dot{\varepsilon} = (A - lc)\varepsilon.
\end{equation}
\end{lemma}

The state estimation is solved if an estimate of $ q $ is obtained. Considering  (\ref{filter1}), the problem to be solved is to estimate the states and unknown input to a minimum phase linear dynamic system.

In order to design the disturbance observer, a reformulation of the system  (\ref{exo}) is first introduced. A controllable pair $ ({F,g}) $ with $ F \in R^{s \times s} $ Hurwitz and $ g \in R^{s} $ are selected. For a matrix $ S $ satisfying A.\ref{A.2} which also implies the pair $ ({S,Q_{(1)}}) $ observable, there exists a non-singular $ M \in R^{s \times s} $ satisfying the following Sylvester equation\cite{bartels1972solution}
\begin{equation*}
MS - FM = g{Q_{(1)}},
\end{equation*}
where $ {Q_{(i)}} $ denotes the $ i $-th row of $ Q $. Let
%\begin{equation*}
$\eta := Mw$ which implies
%\end{equation*}
%we have
\begin{equation}\label{eta}
\dot \eta  = {F_o}\eta ,
\end{equation}
where 
\begin{equation}\label{Fo}
{F_o} = MS{M^{ - 1}} = F + g\psi _1^{\rm{T}},
\end{equation}
and $ \psi _1^{\rm{T}} = {Q_{\left( 1 \right)}}{M^{ - 1}}$. In the new coordinate $ \eta $, $ q $ and $ d_2 $ can be expressed as
\begin{equation}\label{qeta}
q = {\Psi ^{\rm{T}}}\eta,
\end{equation}
where $ {\Psi ^{\rm{T}}} = Q{M^{ - 1}} = {\left[ {{\psi _1},{\psi _2},{\psi _3}, \cdots ,{\psi _{n + 1}}} \right]^{\rm{T}}} $, and 
\begin{equation}\label{deta}
d_2 = \psi _u^{\rm{T}}\eta ,
\end{equation}
with $ \psi _u^{\rm{T}} = {h^{\rm{T}}}{M^{ - 1}} $.

From \eqref{qeta} and \eqref{deta}, we know that if an estimate of $ \eta $ is provided and $ \psi $, $ \psi_u $ are obtained, the estimate of $ q $ and $ d_2 $ are obtained, thus the state estimation is solved. From \eqref{eta} and Lemma \ref{L1}, we have
\begin{equation*}
\dot \eta  = F\eta  + g\left( {\bar y - {p_1} - {\varepsilon _1}} \right),
\end{equation*}
where $ p_1 = p(1) $ and $ \bar{y} = x_1 $, indicating that the observer for $ \eta $ should be designed as
\begin{equation}\label{xi}
\dot \xi  = F\xi  + g\left( {\bar y - {p_1}} \right).
\end{equation}
Define $ e_{\eta}: = \eta - \xi $. We have
%\begin{equation}
${{\dot e}_\eta } = F{e_\eta } - gc\varepsilon$,
%\end{equation}
which together with \eqref{vare} imply
\begin{equation*}
\left[ {\begin{array}{*{20}{c}}
{{{\dot e}_\eta }}\\
{\dot \varepsilon }
\end{array}} \right] = \left[ {\begin{array}{*{20}{c}}
F&{ - gc}\\
0&{A - lc}
\end{array}} \right]\left[ {\begin{array}{*{20}{c}}
{{e_\eta }}\\
\varepsilon 
\end{array}} \right].
\end{equation*}
Since $ F $ and $ A- lc $ are both Hurwitz, $ e_{\eta} $ converges to zero exponentially.

How to get $ \psi $ and $ \psi_u $ depends on whether $ S $ is known or not. From $ q $ and $ d_2 $ in \eqref{qeta} and \eqref{deta} to the observer  (\ref{filter1}), we have
\begin{equation*}
\begin{split}
\psi _i^{\rm{T}}{F_o} & = \psi _{i + 1}^{\rm{T}} - {l_i}\psi _1^{\rm{T}},i = 1, \ldots ,n - 1,\\
\psi _n^{\rm{T}}{F_o} & = \psi _{n + 1}^{\rm{T}} - {l_n}\psi _1^{\rm{T}} + {b_n}\psi _u^{\rm{T}},\\
\psi _{n + 1}^{\rm{T}}{F_o} & =  - {l_{n + 1}}\psi _1^{\rm{T}}.
\end{split}
\end{equation*}
$ S $ is invertible under assumption A.2, relating  (\ref{Fo}) we have $ F_o $ invertible. Then we get
\begin{equation}\label{solvepsi}
\begin{split}
\psi _{i + 1}^{\rm{T}} & = \psi _i^{\rm{T}}{F_o} + {l_i}\psi _1^{\rm{T}},i = 1, \ldots ,n - 1,\\
\psi _{n + 1}^{\rm{T}} & =  - {l_{n + 1}}\psi _1^{\rm{T}}F_o^{ - 1},\\
\psi _u^{\rm{T}} & = \left( {\psi _n^{\rm{T}}{F_o} + {l_n}\psi _1^{\rm{T}} - \psi _{n + 1}^{\rm{T}}} \right)/{b_n}.
\end{split}
\end{equation}
From  (\ref{solvepsi}), we know that if $ \psi_1 $ is obtained, then $ \Psi $ and $ \psi _u $ are obtained. We will show how to get $ \psi _1 $ based on $ S $.

\textbf{Case 1: The matrix $ S $ is entirely known.}

Since $ M $ is non-singular, from  (\ref{Fo}), we know that the matrix $ F_o $ has the same eigenvalues with the matrix $ S $ and then $ \psi _1 $ can be obtained. Without losing generality, $ F $ and $ g $ are selected as
\begin{equation}\label{Fg}
{F = A - b{\alpha _F}}, g = {\left[ {0,0, \cdots ,0,1} \right]^{\rm{T}}}.
\end{equation}
The characteristic polynomial coefficients of $ S $ and $ F $ are 
\begin{equation*}
\begin{split}
{\alpha _S} & = {\left[ {{\alpha _0},{\alpha _1}, \cdots ,{\alpha _{s - 2}},{\alpha _{s - 1}}} \right]^{\rm{T}}},\\
{\alpha _F} & = {\left[ {{f_0},{f_1}, \cdots ,{f_{s - 2}},{f_{s - 1}}} \right]^{\rm{T}}},
\end{split}
\end{equation*}
thus 
\begin{equation}\label{FS}
{\psi _1} = {\alpha _F} - {\alpha _S}.
\end{equation}
Then $ \psi _i,i = 2,\cdots, n+1 $ and $ \psi _u $ are computed from  (\ref{solvepsi}). Therefore, the external disturbance is estimated as
\begin{equation}\label{dm1}
\hat{d_2} = \psi^{\rm{T}}_u \xi.
\end{equation}
To have a summarization, when $ S $ is known, we can get $ \hat d_2 $ with following steps:

\textbf{Procedure 1:}
\begin{itemize}
\item[S1.] Select $ F $ and $ g $ with the form  (\ref{Fg}) and determine $ \alpha _S $ and $ \alpha _F $;
\item[S2.] Compute $ \psi _1 $ using  (\ref{FS}) and $ \psi _i, i = 2,\cdots,n+1 $ and $ \psi _u $ using  (\ref{solvepsi}); 
\item[S3.] Obtain the $ \xi $ using  (\ref{xi});
\item[S4.] Get $ \hat d_2 $ using  (\ref{dm1}).	
\end{itemize}

\textbf{Case 2: The matrix $ S $ is unknown, but $ s $, the dimension of $ S $ is known}

In this case, we know that $ \psi _1 \in  R^{s} $. Since $ S $ is unknown, we cannot obtain $ \psi _1 $ through \textbf{Procedure 1}. Suppose that $ \hat{\psi} _1 $ is the estimate of $  {\psi_1} $, then $ \zeta $, the estimate of $ \xi $ is updated by
\begin{equation}\label{zeta}
\dot \zeta  = F\zeta  + g\hat \psi _1^{\rm{T}}\xi.
\end{equation}
Define $ e_{\xi} = \xi - \zeta $, we have
\begin{equation*}
{{\dot e}_\xi } = F{e_\xi } + gc\varepsilon  + g\psi _1^{\rm{T}}{e_\eta } + g{\xi ^{\rm{T}}}{{\tilde \psi }_1}.
\end{equation*}
Define $ e = {\left[ {\begin{array}{*{20}{c}}
{e_\xi ^{\rm{T}}}&{e_\eta ^{\rm{T}}}&{{\varepsilon ^{\rm{T}}}}
\end{array}} \right]^{\rm{T}}} $, we have
\begin{equation*}
\dot e = {A_c}e + {\phi}\left( t \right){{\tilde \psi }_1},
\end{equation*}
where $ {A_c} = \left[ {\begin{array}{*{20}{c}}
F&{g\psi _1^{\rm{T}}}&{gc}\\
0&F&{ - gc}\\
0&0&{A - lc}
\end{array}} \right] $ and $ \phi \left( t \right) = \left[ {\begin{array}{*{20}{c}}
{g{\xi ^{\rm{T}}}}\\
0\\
0
\end{array}} \right] $. $ A_c $ is Hurwitz for that both of $ F $ and $ A-lc $ are Hurwitz, therefore,  for a given positive definite symmetric matrix $ Q_c $, there exists a positive definite symmetric matrix $ P_c $ satisfying the Lyapunov equation
\begin{equation*}
A_c^{\rm{T}}{P_c} + {P_c}{A_c} =  - 2{Q_c}.
\end{equation*}
Selecting $ \Gamma \in R^{s \times s} $ as a positive definite matrix, the  Lyapunov candidate function is selected as
\begin{equation*}
V\left( {e,{{\tilde \psi }_1}} \right) = \frac{1}{2}\left( {{e^{\rm{T}}}{P_c}e + \tilde \psi _1^{\rm{T}}{\Gamma ^{ - 1}}{{\tilde \psi }_1}} \right),
\end{equation*} 
whose first derivative is
\begin{equation*}
\dot V\left( {e,{{\tilde \psi }_1}} \right)  =  - {e^{\rm{T}}}{Q_c}e + \tilde \psi _1^{\rm{T}}\phi \left( t \right){P_c}e + \tilde \psi _1^{\rm{T}}{\Gamma ^{ - 1}}{{\dot {\tilde \psi }}_1},
\end{equation*}
by setting
\begin{equation*}
\tilde \psi _1^{\rm{T}}\phi^{T} \left( t \right){P_c}e + \tilde \psi _1^{\rm{T}}{\Gamma ^{ - 1}}{{\dot {\tilde \psi} }_1} = 0,
\end{equation*}
which indicates that
\begin{equation}\label{updatelaw1}
{{\dot {\tilde \psi} }_1} =  - {\Gamma ^{ - 1}}\phi ^{\rm{T}} \left( t \right){P_c}e,
\end{equation}
we have
\begin{equation*}
\dot V\left( {e,{{\tilde \psi }_1}} \right) =  - {e^{\rm{T}}}{Q_c}e, Q_c >0,
\end{equation*}
so $ e $ and $ {\tilde \psi} _1 $ is bounded and from the well known Barbalat Lemma we know that $ \mathop {\lim }\limits_{t \to \infty } e\left( t \right) = 0 $. 

Since $ {\psi }_1  $ is unknown, we cannot get $ P_c $, and the updating law  (\ref{updatelaw1}) is not implementable. Suppose that $ P_c $ is of the form
\begin{equation*}
{P_c} = diag\{ {P_1},{\gamma _1}{P_1},{\gamma _2}{P_2}\},
\end{equation*}
where $ P_1 $ and $ P_2 $ are positive definite matrices satisfying
\begin{equation}\label{lyapQ1}
{F^{\rm{T}}}{P_1} + {P_1}F =  - 2{Q_1},
\end{equation}
and
\begin{equation*}
{\left( {A - lc} \right)^{\rm{T}}}{P_2} + {P_2}\left( {A - lc} \right) =  - 2{Q_2},
\end{equation*}
with $ Q_1 $ and $ Q_2 $ selecting as positive definite matrices and $ \gamma _1 $ and $ \gamma _2 $ are positive constant. Thus $ Q_c $ can be selected as
\begin{equation*}
{Q_c} = \left[ {\begin{array}{*{20}{c}}
{{Q_1}}&{\frac{{{P_1}g\psi _1^{\rm{T}}}}{{ - 2}}}&{\frac{{{P_1}gc}}{{ - 2}}}\\
{\frac{{{\psi _1}{g^{\rm{T}}}{P_1}}}{{ - 2}}}&{{\gamma _1}{Q_1}}&{\frac{{{\gamma _1}{P_1}gc}}{2}}\\
{\frac{{{c^{\rm{T}}}{g^{\rm{T}}}{P_1}}}{{ - 2}}}&{\frac{{{\gamma _1}{c^{\rm{T}}}{g^{\rm{T}}}{P_1}}}{2}}&{{\gamma _2}{Q_2}}
\end{array}} \right].
\end{equation*}
Obviously, $ Q_c $ is symmetric and by selecting $ \gamma_1 $ and $ \gamma_2 $ sufficiently large, $ Q_c $ will be positive definite. Then the updating law  (\ref{updatelaw1}) can be rewritten as
\begin{equation}\label{updatelaw2}
{{\dot {\tilde \psi }}_1} =  - {\Gamma ^{ - 1}}\xi {g^{\rm{T}}}{P_1}{e_\xi },
\end{equation}
yielding
\begin{equation}\label{updatelaw3}
{{\dot {\hat \psi }}_1} = {\Gamma ^{ - 1}}\xi {g^{\rm{T}}}{P_1}{e_\xi }.
\end{equation}
 (\ref{updatelaw3}) is implementable for that $ \Gamma $ and $ g $ are selected, $ P_1 $ is computed by  (\ref{lyapQ1}), $ \xi $ is updated by  (\ref{xi}), $ e_{\xi} = \xi - \zeta $ where $ \zeta $ is updated by  (\ref{zeta}).

The updating law  (\ref{updatelaw3}) ensures that $ \mathop {\lim }\limits_{t \to \infty } {{\dot {\tilde \psi} }_1} = 0 $, indicating that $ {{\tilde \psi }_1}  $ will converge to a constant vector, but no guarantee that $ {{\tilde \psi }_1}  $ converges to zero. It can be proven that $ {{\tilde \psi }_1}  $ converges to zero iff $ \xi g^{\rm{T}} $ is persistently excited. Computing
\begin{equation*}
\int_{{t_0}}^{{t_0} + {T_0}} {\xi {g^{\rm{T}}}{{\left( {\xi {g^{\rm{T}}}} \right)}^{\rm{T}}}d\tau }  = {\left\| g \right\|^2}\int_{{t_0}}^{{t_0} + {T_0}} {\xi {\xi ^{\rm{T}}}d\tau },
\end{equation*}
where $ \xi =  {\xi \left( \tau  \right) =  \eta (\tau) - e_\eta (\tau) } $, we have $ \xi g^{\rm{T}} $ is persistently excited iff $ \eta $ is persistently excited for that $ e_{\eta} $ converges to zero exponentially. Since
\begin{equation*}
\int_{{t_0}}^{{t_0} + {T_0}} {\eta {\eta ^{\rm{T}}}d\tau }  = {\left\| M \right\|^2}\int_{{t_0}}^{{t_0} + {T_0}} {w{w^{\rm{T}}}d\tau },
\end{equation*}
we have $ \eta $ is persistently excited iff $ w $ is persistently excited, which can be realized by selecting a proper $ w_0 $.

With the estimate of $ \psi _1 $, we obtain the estimate of $ F_o $ as
\begin{equation*}
{{\hat F}_o} = F + g\hat \psi _1^{\rm{T}},
\end{equation*}
and the estimate of $ \psi _i, i = 2,\cdots, n+1 $ and $ \psi _u $  as
\begin{equation}\label{Psim}
\begin{split}
\hat \psi _{i + 1}^{\rm{T}} & = \hat \psi _i^{\rm{T}}{{\hat F}_o} + {l_i}\hat \psi _1^{\rm{T}},i = 1, \ldots ,n - 1,\\
\hat \psi _{n + 1}^{\rm{T}} & =  - {l_{n + 1}}\hat \psi _1^{\rm{T}}\hat F_o^{ - 1},\\
\hat \psi _u^{\rm{T}} & = \left( {\hat \psi _n^{\rm{T}}{{\hat F}_o} + {l_n}\hat \psi _1^{\rm{T}} - \hat \psi _{n + 1}^{\rm{T}}} \right)/{b_n}.
\end{split}
\end{equation}
Spontaneously we get the estimate of the external disturbance
\begin{equation}\label{dm2}
\hat{d_2} = \hat \psi _u^{\rm{T}}\xi.
\end{equation}
To have a summarization, when $ S $ is unknown, we can get $ \hat d_2 $ with following steps:

\textbf{Procedure 2:}
\begin{itemize}
\item[S1.] Select $ F $ and $ g $ with the form  (\ref{Fg}) and $ Q_1 $, then compute $ P_1 $ from  (\ref{lyapQ1});
\item[S2.] Obtain $ \xi $ using  (\ref{xi});
\item[S3.] Obtain $ \zeta $ using  (\ref{zeta});
\item[S4.] Update $ \hat \psi _1 $ using  (\ref{updatelaw3}) ;
\item[S5.] Compute $ \hat \psi _i, i = 2,\cdots,n+1 $ and $ \hat \psi _u $ using  (\ref{Psim}); 
\item[S6.] Get $ \hat d_2 $ using  (\ref{dm2}).	
\end{itemize}

\section{The internal model based active disturbance rejection control design and analysis}\label{IADRC}
\begin{theorem}
Considering the dynamic system  (\ref{sys2}) satisfying assumption A.\ref{A.1} and the following output feedback observer
\begin{equation}\label{eso1}
\left\{ \begin{array}{l}
	\dot v = Av + bu + l\left( {\bar y - y} \right)\\
	y = cv\\	
\end{array} \right.,
\end{equation}
and the control input
\begin{equation}\label{adrc}
u = -k^{T}v,
\end{equation}
the closed-loop system described under the state $ z = {\left[ {{x^{\rm{T}}},{v^{\rm{T}}}} \right]^{\rm{T}}} $ is asymptotically stable when $ l $ is selected such that  $ A - lc $ is Hurwitz and $ k $ is selected as $ k = {\left[ {{{\bar k}^{\rm{T}}},1} \right]^{\rm{T}}} $ where $ \bar{k} $ is selected such that $ \bar{A} - \bar{b}\bar{k} $ is Hurwitz.

\end{theorem}

\begin{remark}
In fact,  (\ref{eso1}) and  (\ref{adrc}) are the BADRC for system  (\ref{sys2}).
\end{remark}

With the estimate of the external disturbance, the controller is then designed as
\begin{equation*}
u = {u_c} + {u_d},
\end{equation*}
where 
\begin{equation*}
u_d = -\hat{d_2},
\end{equation*}
and $ u_c $ is generated by
\begin{equation}\label{meso}
\begin{array}{l}
\dot v = Av + b{u_c} + l\left( {\bar y - cv} \right),\\
{u_c} =  - k^{T}v.
\end{array}
\end{equation}

\begin{remark}
In  (\ref{meso}), the input to get $ v $ is $ u_c $ rather than $ u $, which is reasonable for that the external disturbance $ d_2 $ is compensated and $ u_c $ can be seen as the feedback control when there is no disturbance.
\end{remark}

\textbf{Closed-loop system stability analysis:}
We consider the stability of the original system  (\ref{sys2}) under the control  (\ref{meso}). Defining $ {\tilde v} = x - v $, we have
\begin{equation*}
\dot {\tilde v} = \left( {A - lc} \right)\tilde v + b{{\tilde d}_2} + f,
\end{equation*}
and
\begin{equation*}
\dot x = \left( {A - bk} \right)x + bk\tilde v + b{{\tilde d}_2} + f,
\end{equation*}
which together imply
\begin{equation*}
\begin{split}
\left[ {\begin{array}{*{20}{c}}
	{\dot x}\\
	{\dot {\tilde v}}
\end{array}} \right] = & \left[ {\begin{array}{*{20}{c}}
{A - bk}&{bk}\\
0&{A - lc}
\end{array}} \right]\left[ {\begin{array}{*{20}{c}}
x\\
{\tilde v}
\end{array}} \right] \\
& + \left[ {\begin{array}{*{20}{c}}
b&{{b_f}}\\
b&{{b_f}}
\end{array}} \right]\left[ {\begin{array}{*{20}{c}}
{{{\tilde d}_2}}\\
{{{\dot x}_{n + 1}}}
\end{array}} \right].
\end{split}
\end{equation*}
Since $ \tilde{d}_2 $ and $ \dot{x}_{n+1} $ are bounded, the overall system is input-to-state-stable (ISS).

\section{Simulation examples}\label{simulationexamples}
Consider the following system
\begin{equation*}
\left\{ \begin{array}{l}
{{\dot x}_1} = {x_2}\\
{{\dot x}_2} = {f_2}\left( x,\omega _1 \right) + {b_2}\left( {u + \omega_2} \right)\\
y = {x_1}
\end{array} \right.,
\end{equation*}
where $ b_2 = 3 $ is a known constant, $ {f_2}\left( x,\omega _1 \right) = 0 $ and $ \omega _2 = \sigma _0 + r\sin(\sigma t+ \varphi)$. So $ d_1 = {\sigma _0}/{b_2} $, and $ d_2 $ is of the form
\begin{equation*}
d_2 = r\sin(\sigma t+ \varphi),
\end{equation*}
where $ r = 0.8 $, $ \sigma = 2 $ and $ \varphi = {\pi}/{5} $. $ d_2 $ can be rewritten as  (\ref{exo}) where $ h = {\left[ {r\cos \varphi ,r\sin \varphi } \right]^{\rm{T}}} $, and $ w = {\left[ {\sin \sigma t,\cos \sigma t} \right]^{\rm{T}}} $ with $ S = \left[ {\begin{array}{*{20}{c}}
0&\sigma \\
{ - \sigma }&0
\end{array}} \right]  $ and $ {w_0} = {\left[ {0,1} \right]^{\rm{T}}} $.

\begin{figure}
	\begin{minipage}[t]{0.5\linewidth}
	\begin{center}
		\includegraphics[height=0.65\textwidth]{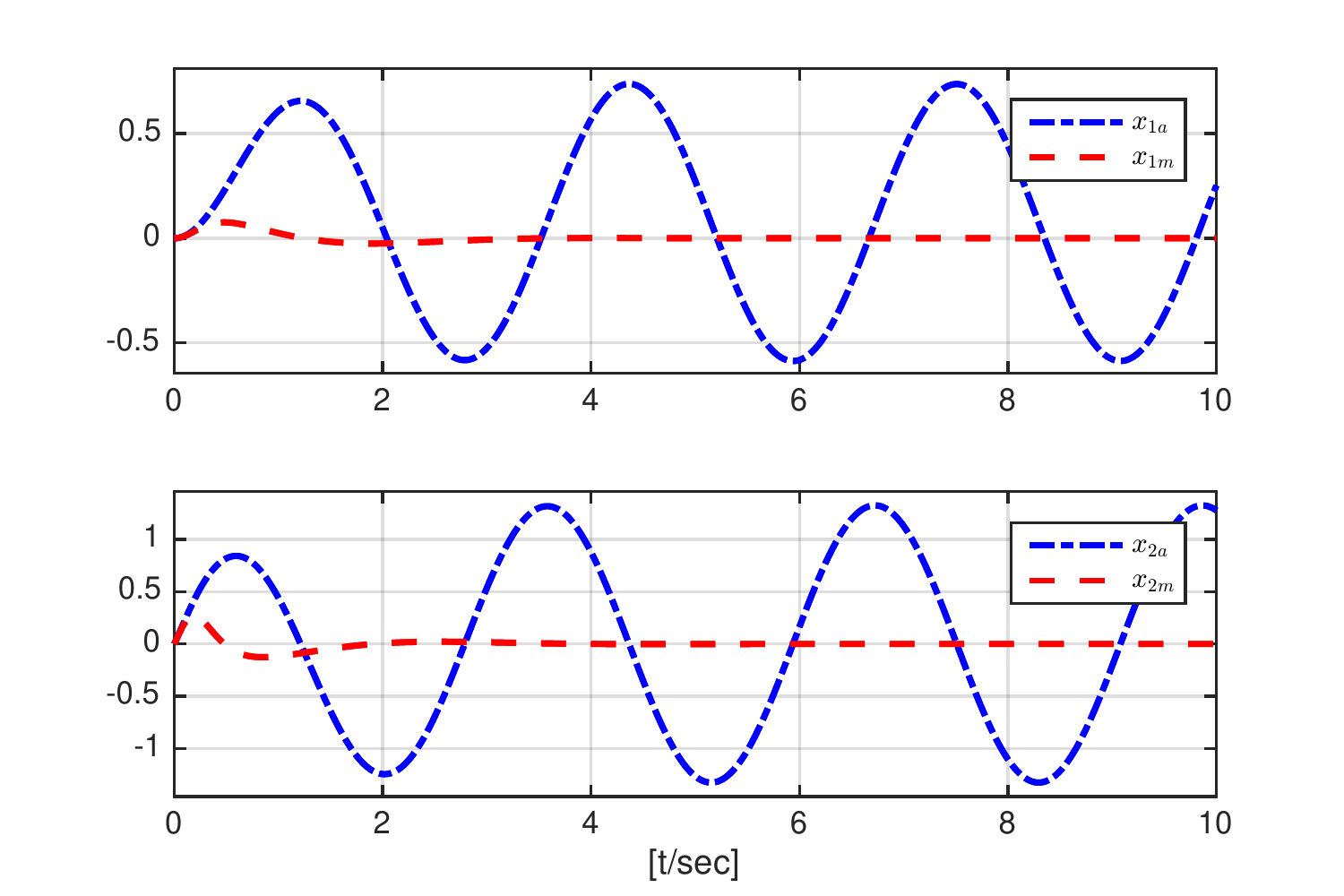}
		\caption{States revolution w.r.t ADRC  and IADRC (known $ S $)}\label{fig_ss}
	\end{center}
	\end{minipage}%
	\begin{minipage}[t]{0.5\linewidth}
		\begin{center}
			\includegraphics[height=0.65\textwidth]{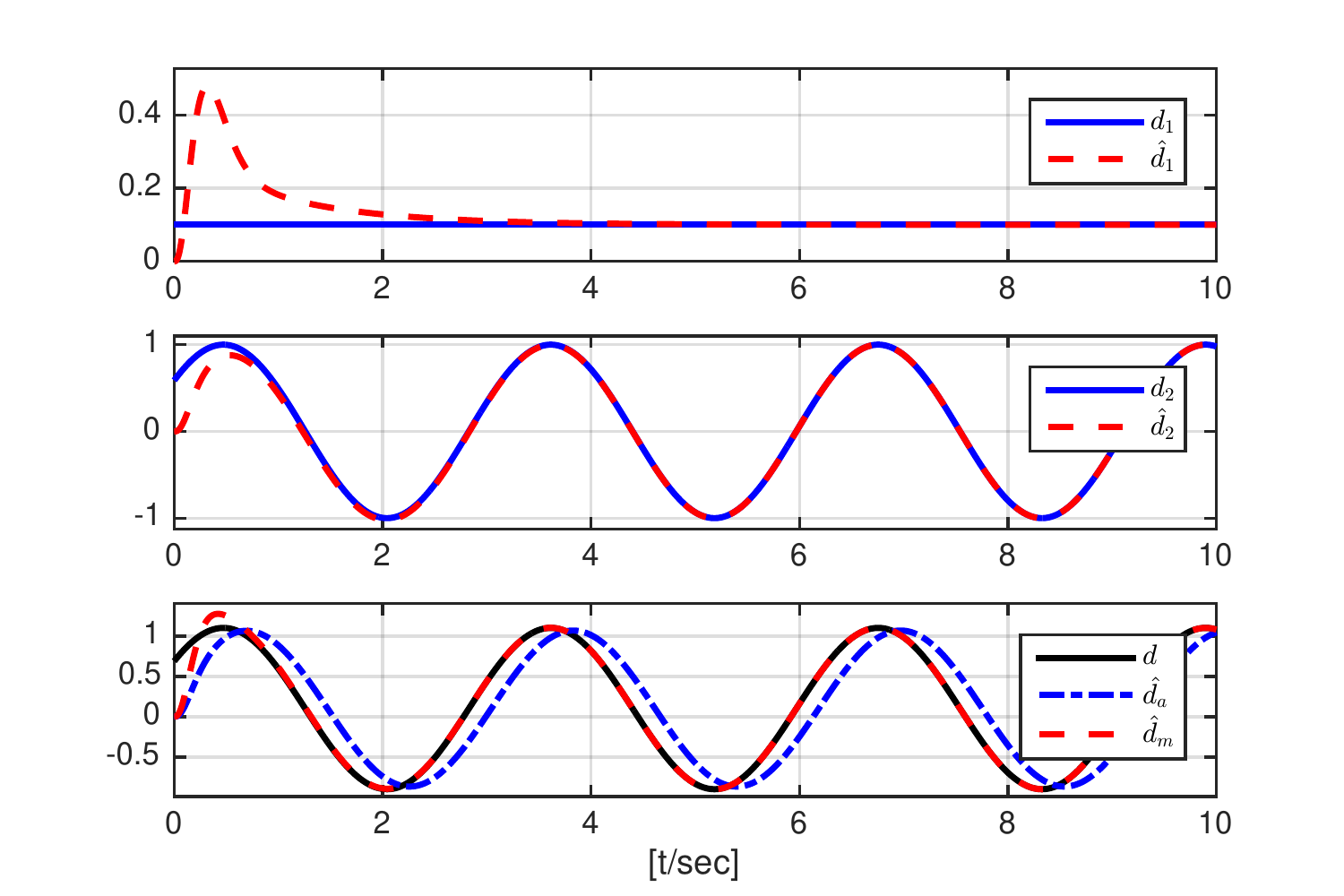}
			\caption{Disturbances and their estimates(known $ S $)}\label{fig_sd}
		\end{center}
	\end{minipage}
\end{figure}

We first consider the case $ S $ is known. With \textbf{Procedure 1}, we select $ F = \left[ {\begin{array}{*{20}{c}}
0&1\\
{ - 2}&{ - 3}
\end{array}} \right] $, $ g = {\left[ {0,1} \right]^{\rm{T}}} $, the we get $ {F_o} = \left[ {\begin{array}{*{20}{c}}
0&1\\
{ - 4}&0
\end{array}} \right] $, $ {\psi _1} = {\left[ {-2,3} \right]^{\rm{T}}} $, $ {\psi _2} = {\left[ {-102,133} \right]^{\rm{T}}} $, $ {\psi _3} = {\left[ {900,1500} \right]^{\rm{T}}} $, and $ {\psi _u} = {\left[ {7168/3,1116} \right]^{\rm{T}}} $. The simulation results are shown in Figs.\ref{fig_ss} and \ref{fig_sd}. In Fig.\ref{fig_ss}, $ x_{1a} $ and $ x_{2a} $ are states with respect to (w.r.t) the BADRC while $ x_{1m} $ and $ x_{2m} $ are states w.r.t the proposed IADRC. Fig.\ref{fig_ss} shows that the proposed algorithm can reject the unknown disturbance more effective than the BADRC. Fig.\ref{fig_sd} reveals the reason: by exploiting the known information of the disturbance, the perfect estimation of $ d_2 $ is achieved while there exists a phase lag when the disturbance is estimated only by using ESO.

\begin{figure}
	\begin{minipage}[t]{0.5\linewidth}
		\begin{center}
			\includegraphics[height=0.65\textwidth]{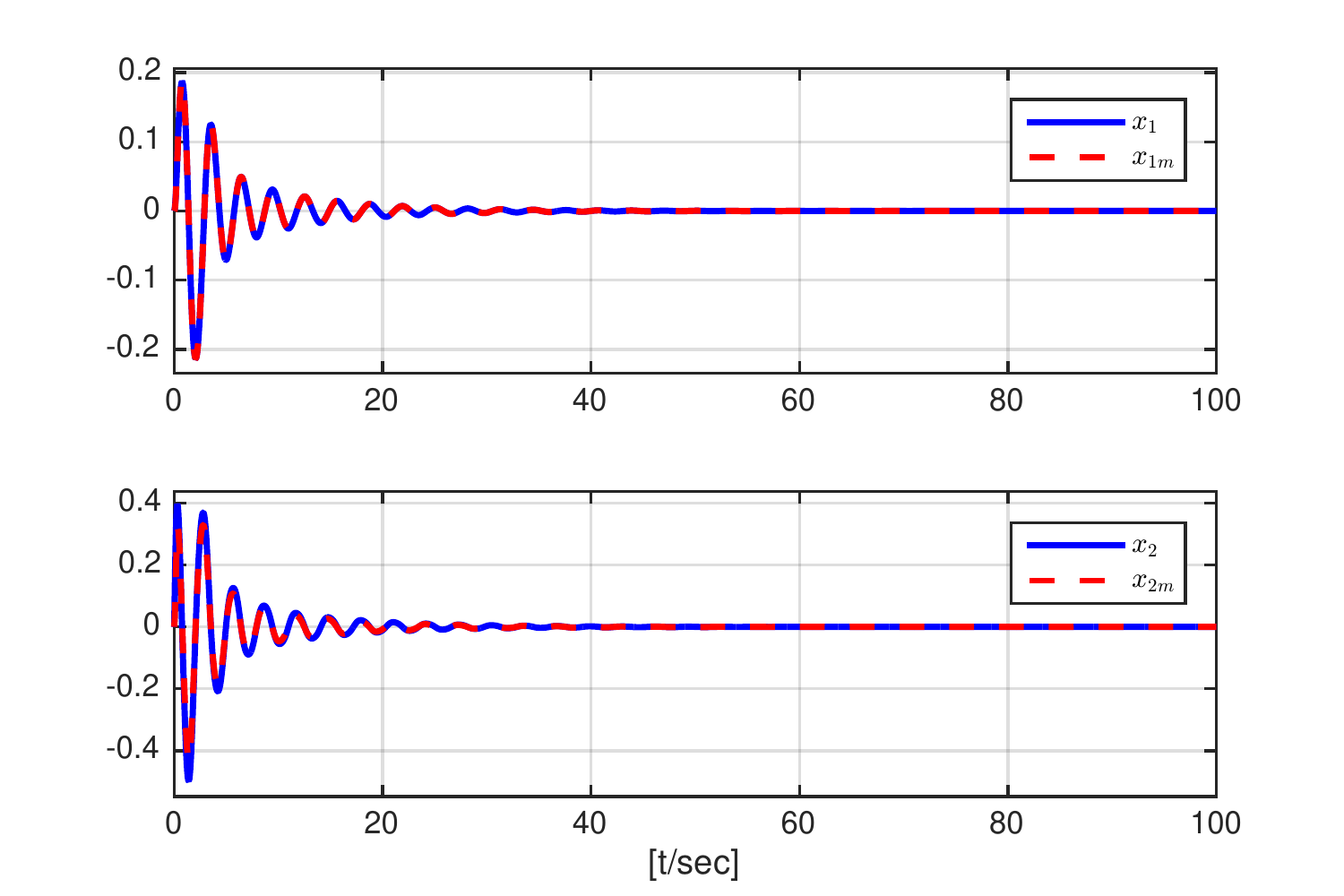}
			\caption{States revolution w.r.t IADRC (unknown $ S $)}\label{fig_us}
\end{center}
	\end{minipage}%
	\begin{minipage}[t]{0.5\linewidth}
		\begin{center}
		\includegraphics[height=0.65\textwidth]{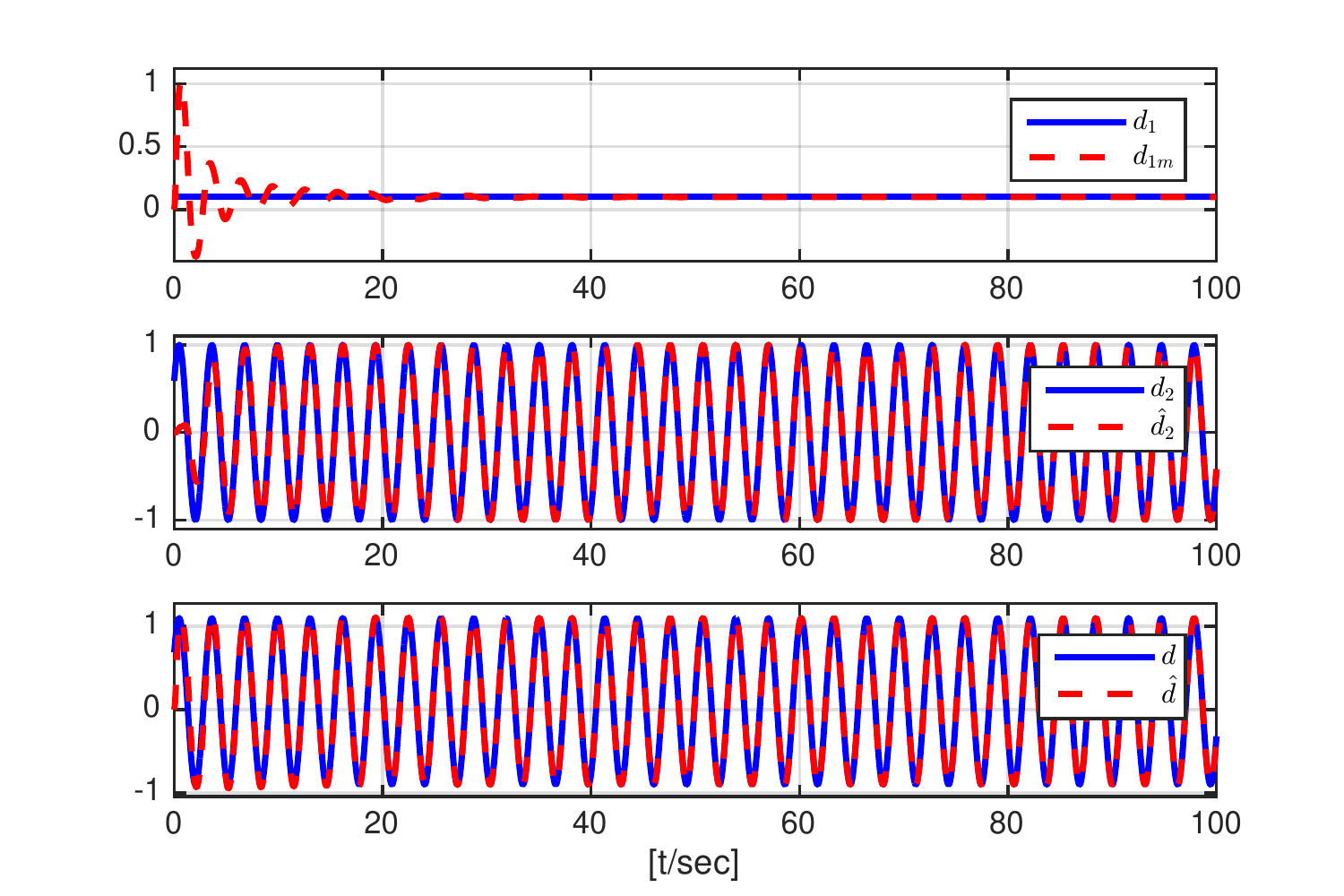}
		\caption{Disturbances and their estimates (unknown $ S $)}\label{fig_ud}
\end{center}
	\end{minipage}
\end{figure}

\begin{figure}
	\begin{minipage}[t]{0.5\linewidth}
		\begin{center}
			\includegraphics[height=0.65\textwidth]{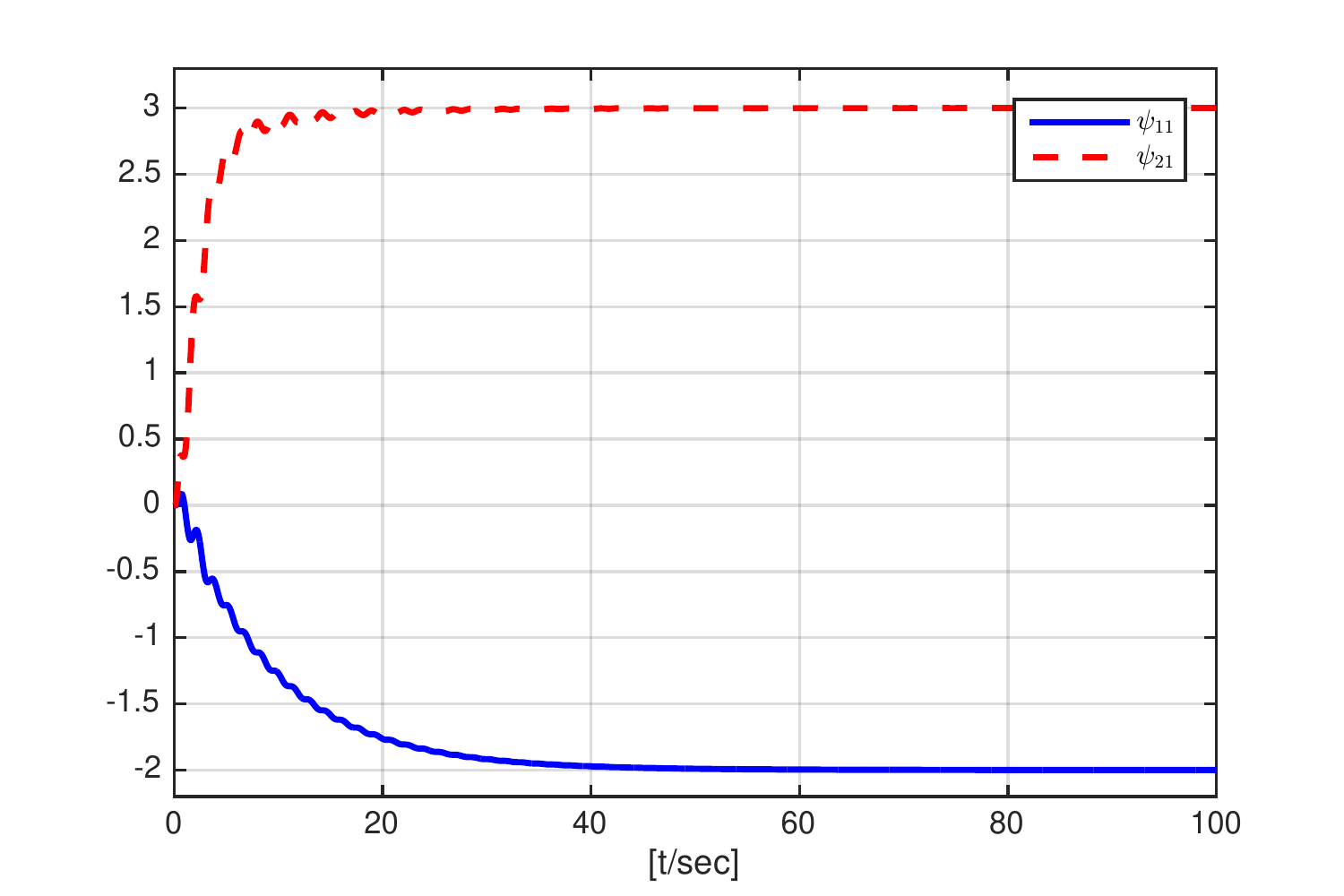}
			\caption{Estimates of $ \psi _{11} $ and $ \psi _{21} $}\label{fig_ukesa1}
		\end{center}
	\end{minipage}%
	\begin{minipage}[t]{0.5\linewidth}
		\begin{center}
			\includegraphics[height=0.65\textwidth]{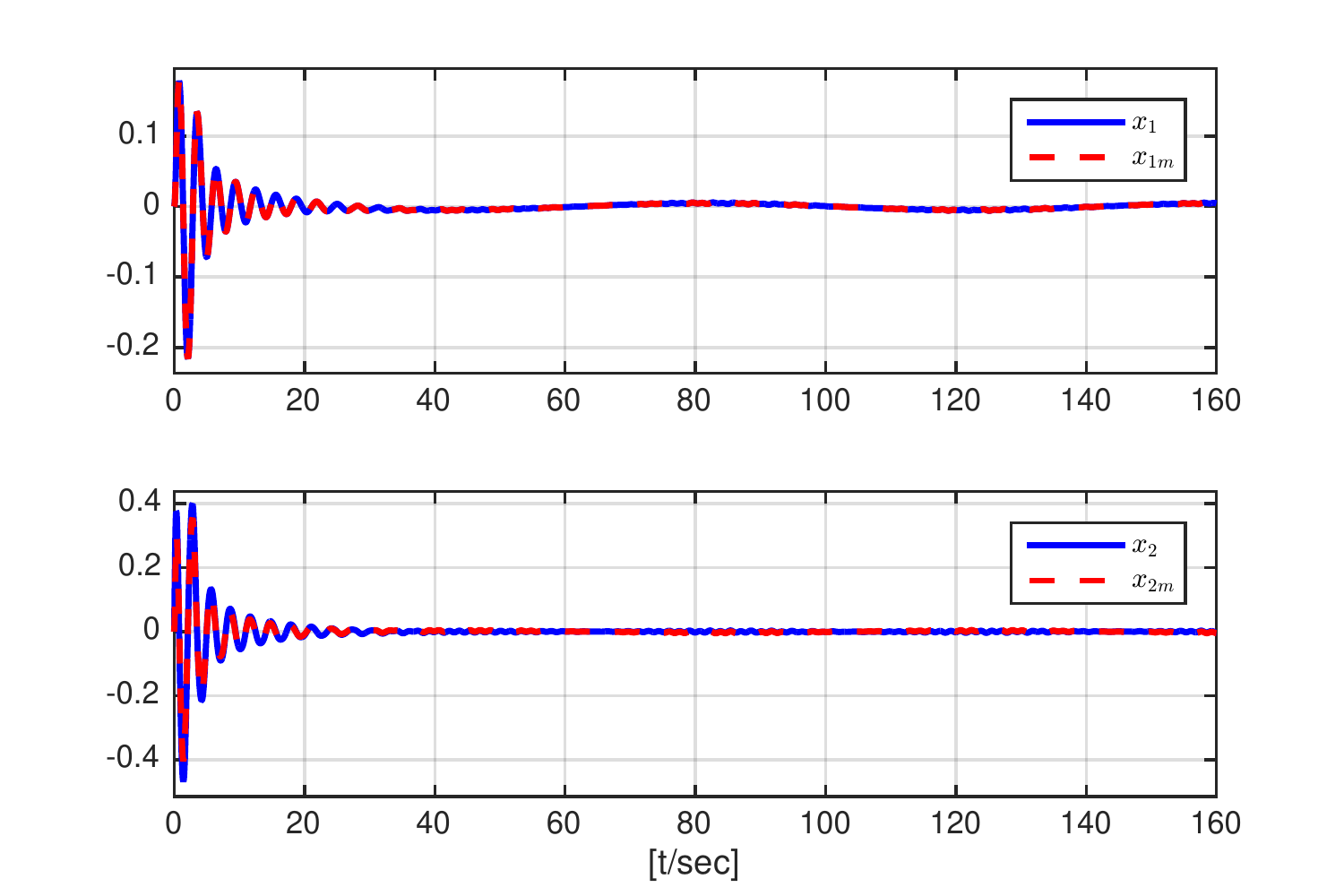}
			\caption{States revolution w.r.t IADRC (unknown $ S $)}\label{fig_ucs}
		\end{center}
	\end{minipage}
\end{figure}

We then consider the case $ S $ is unknown. With \textbf{Procedure 2}, we select $ F = \left[ {\begin{array}{*{20}{c}}
0&1\\
{ - 2}&{ - 3}
\end{array}} \right] $, and $ g = {\left[ {0,1} \right]^{\rm{T}}} $, $ \Gamma  = 40000{{\bf{I}}_2} $ and $ {Q_1} = 150{{\bf{I}}_2} $. By solving Lyapunov equation  (\ref{lyapQ1}), we have $ {P_1} = \left[ {\begin{array}{*{20}{c}}
{300}&{ - 150}\\
{ - 150}&{150}
\end{array}} \right] $. Results are shown in Figs.\ref{fig_us}, \ref{fig_ud} and \ref{fig_ukesa1}. As shown in Fig.\ref{fig_ud}, perfect estimates for $ d_1 $ and $ d_2 $ are achieved for that the estimate of $ \psi _1 $ converges to its real value $ {\left[ {-2,3} \right]^{\rm{T}}}$ as shown in Fig.\ref{fig_ukesa1}. The total matched disturbance $ d $ is perfectly estimated thus being fully compensated, therefore the system states converges to zero at the steady state shown in Fig.\ref{fig_us}.

\begin{figure}
	\begin{minipage}[t]{0.5\linewidth}
		\begin{center}
			\includegraphics[height=0.65\textwidth]{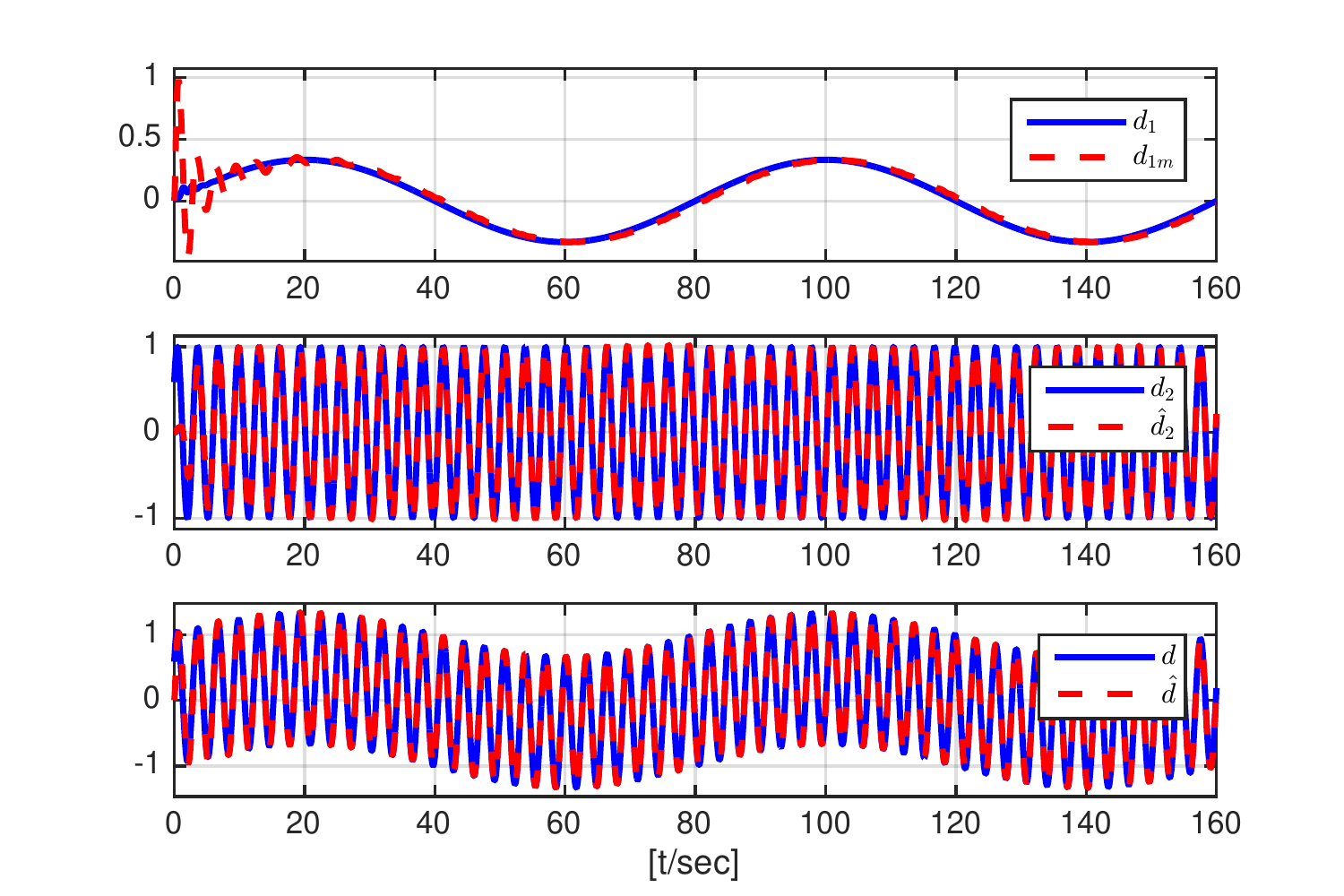}
			\caption{Disturbances and their estimates (unknown S)}\label{fig_ucd}
		\end{center}
	\end{minipage}%
	\begin{minipage}[t]{0.5\linewidth}
		\begin{center}
			\includegraphics[height=0.65\textwidth]{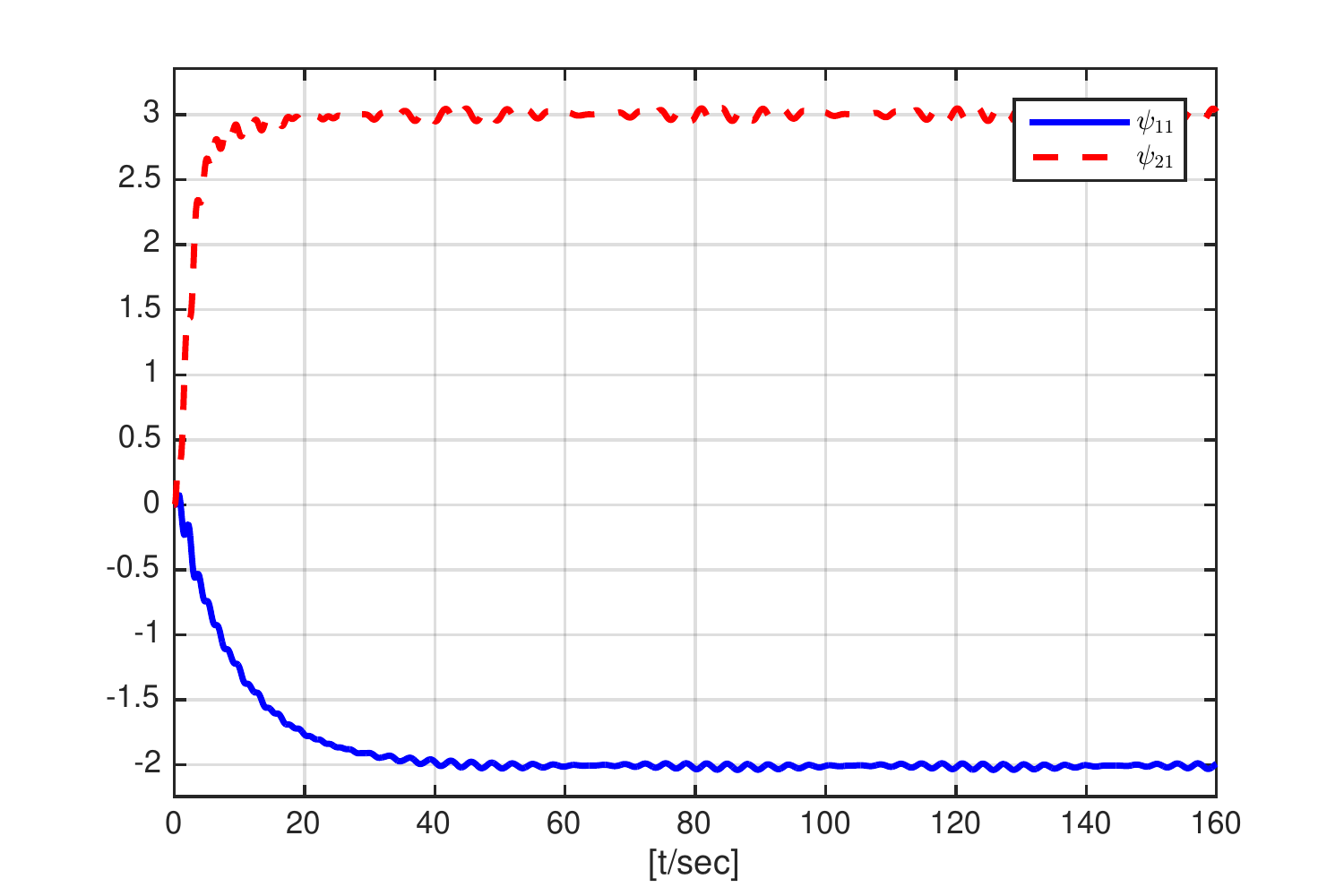}
			\caption{Estimates of $ \psi _{11} $ and $ \psi _{21} $ (unknown S)}\label{fig_uckesa1}
		\end{center}
	\end{minipage}
\end{figure}

A more complex case is considered here. Suppose that $ f_2(x,\omega_1) = {x^2_1} + {x^2_2} + \sin(\frac{\pi}{40}t) $, therefore $ d_1 = {f_2}/{b_2} + \sigma_0 $ and $ d_2 = r\sin(\sigma t+ \varphi) $. $ S $ is unknown and parameters are chosen the same as above. Results are shown in Figs.\ref{fig_ucs}, \ref{fig_ucd} and \ref{fig_uckesa1}. Since $ x_3 = f_2 + {b_2} \sigma _0 $ is not a constant, as shown in Fig.\ref{fig_ucd} no perfect tracking for $ d_1 $ can be reached, which leads to the estimates of $ {\psi _1} $ oscillating around its true value in a small region as shown in Fig.\ref{fig_uckesa1}. Therefore, there exists oscillation in system states around $ 0 $ as shown in Fig.\ref{fig_ucs}.

\section{Conclusions}\label{conclusions}
The principle of ADRC from the internal model principle point of view was presented in this paper.  An improved ADRC that can properly exploit known information about the disturbance was proposed. Depending on whether the dynamics of $S$ is known or not two adaptation algorithms were provided. Moreover, it was shown that when $ S $ is unknown, it is required to estimate it, whereas the system states are the only elements that need to be estimated when $S$ is known. Simulation results show that IADRC is of significant improvement compared to the BADRC.

%%%%%%%%%%%%%%%%%%%%%%%%%%%%%%%%%%%%%%%%%%%%%%%%%%%%%%%%%%%%%%%%%%%%%%%%%%%%%%%%

\section*{ACKNOWLEDGMENT}
This work was supported by \textit{CAS}, \textit{USTC Special Grants for Postgraduate Research, Innovation and Practice}, and \textit{USTC Youth Innovation Fund WK2100100016}.

%%%%%%%%%%%%%%%%%%%%%%%%%%%%%%%%%%%%%%%%%%%%%%%%%%%%%%%%%%%%%%%%%%%%%%%%%%%%%%%%

\bibliographystyle{unsrt}
\bibliography{ACC_2016_FINAL}

%\addtolength{\textheight}{-12cm}   % This command serves to balance the column lengths
% on the last page of the document manually. It shortens
% the textheight of the last page by a suitable amount.
% This command does not take effect until the next page
% so it should come on the page before the last. Make
% sure that you do not shorten the textheight too much.

\end{document}